\documentclass[aip,jcp,twocolumn,showpacs,superscriptaddress,amsmath,amssymb,floatfix,10pt]{revtex4-1}  
\usepackage{graphicx}  
\usepackage{dcolumn}   
\usepackage{bm}        
\usepackage{amssymb}   
\usepackage{framed}
\usepackage{amsmath}
\usepackage{multirow}
\usepackage{hhline}
\usepackage{color}
\definecolor{bluegreen}{rgb}{0,0.2,0.8}
\usepackage{subfigure,amsmath,verbatim,moreverb}
\usepackage{tabularx}
\usepackage{adjustbox}
\usepackage{lipsum}

\begin{document}

\widetext
\title{Inhomogeneity Induced and Appropriately Parameterized Semilocal Exchange and Correlation
Energy Functionals in Two-Dimensions}

\author{Abhilash Patra}
\email{abhilashpatra@niser.ac.in}
\affiliation{School of Physical Sciences, National Institute of Science Education and Research, HBNI, Bhubaneswar 752050, India}
\author{Subrata Jana}
\affiliation{School of Physical Sciences, National Institute of Science Education and Research, HBNI, Bhubaneswar 752050, India}
\author{Prasanjit Samal}
\email{psamal@niser.ac.in}
\affiliation{School of Physical Sciences, National Institute of Science Education and Research, HBNI, Bhubaneswar 752050, India}

\date{\today}

\begin{abstract}

The construction of meta generalized gradient approximations based on the density matrix expansion~(DME)
is considered as one of the most accurate technique to design semilocal exchange energy functionals in
two-dimensional density functional formalism. The exchange holes modeled using DME possess unique
features that make it a superior entity. Parameterized semilocal exchange energy functionals based on the 
DME are proposed. The use of different forms of the momentum and flexible parameters is to subsume the 
non-uniform effects of the density in the newly constructed semilocal functionals. In addition to the 
exchange functionals, a suitable correlation functional is also constructed by working upon the local 
correlation functional developed for 2D homogeneous electron gas~(2D-HEG). The non-local effects are 
induced into the correlation functional by a parametric form of one of the newly constructed exchange energy 
functionals. The proposed functionals are applied to the parabolic quantum dots with a varying number of 
confined electrons and the confinement strength. The results obtained with the aforementioned functionals 
are quite satisfactory which indicates why these are suitable for two-dimensional quantum systems.

\end{abstract}

\maketitle

\section{Introduction}
The Density functional theory~(DFT) is a most favored formalism~\cite{hk64,ks65} in condensed matter physics 
and quantum chemistry for electronic structure calculations. The Kohn-Sham (KS) formalism is the mainstay of 
DFT which solves an auxiliary one particle Schr\"{o}dinger like equations and provides ground state energy 
and other properties as a functional of density and density derived quantities. In density functional formalism 
the only unknown part is the exchange-correlation $(xc)$ functional, which is a small fraction but most important 
part of the total energy. So, several techniques and approximations are introduced which play crucial role in 
designing the non-trivial entity of DFT with different formal properties. Therefore, precise approximations 
to XC energy functional ($E_{xc}$) are always been an exciting and enthralling research field. The first ever 
XC functional construction is the spin-polarized approximation from uniform electron gas limit called the local 
spin-density approximation~(LSDA) i.e. the $E_{xc}$ is a function of only electron density. Then inhomogeneity 
is added to this functional through the gradient of density and kinetic energy density, which gave rise to the 
generalized gradient approximations (GGAs) and meta-GGAs~\cite{b83,pw86,b88,pbe96,br89,pkzb,tpss,tamo,nv1,kos,
vs97,vsxc98,tsuneda}. All these functionals are developed for three-dimensional~(3D) systems. Previous studies 
show that the semilocal density functional proposed for 3D systems cannot be applied directly to the~(pure) 
two-dimensional (2D) systems due to the dimensional crossover problem~\cite{klnl,lpjp,lcjp}. 

Over the last few decades, increasing attention has been paid to the density functional study of the low
dimensional quantum systems, which includes semiconductor quantum dots, quantum point contacts, and quantum 
Hall systems. For this, meticulous developments of $E_{xc}$ in 2D is very essential. The 2D-LDA~\cite{rk} 
seems to be a valuable option instead of the 3D-LDA for accurately studying pure 2D systems in a pure 2D grid. 
So, within DFT formalism, quantum dots~(QDs) are commonly studied using 2D-LSDA exchange along with locally 
approximated correlation~\cite{attaccalite}. But, to take care of the inhomogeneity present in the system, 
further improvements over 2D-LSDA, such as 2D-GGAs and 2D-meta-GGAs are also proposed in recent years
~\cite{prvm,prhg,prg,pr1,prp,rp,sr,pr2,rpvm,rpcp9,prlvm,vrmp}.

In this present work, our focus is on the development of semilocal exchange energy functionals using the 
advanced DME techniques as an intellectual basis. Then, we will propose an appropriate correlation energy 
functional using one of the semilocal exchange energy functionals designed for the 2D systems. Thus, as a 
first step, three exchange energy functionals are constructed with four adjustable parameters each. The 
newly constructed semilocal exchange energy functionals depend on the gradient of density and kinetic energy 
density. The inhomogeneity associated with the system is imposed on the functional construction through the 
modification of the Fermi momentum. Thus, the Fermi momentum is modified by adding appropriate functional 
forms involving the reduced density gradient and kinetic energy density with suitable remodeling. These 
transformed momenta are used in the newly developed exchange energy functionals. Since, the popularly used 
2D-LDA correlation energy~\cite{attaccalite} was constructed for 2D electron gas, by fitting with Monte-Carlo 
simulation data and including low- and high-density limits. So, to construct the correlation functional 
compatible with the above exchange functionals, we have modified the 2D-LDA correlation functional by 
imposing the non-local effects through an exchange energy functional using appropriate parameters. The 
numerical investigations of these newly constructed semilocal functionals are done by applying these to
parabolic QD systems. The calculations are done by varying the number of confined electrons and confinement 
strength of the parabolic quantum dots. We have compared our results with the 2D exact exchange~(2D-EXX) 
within the Krieger-Li-Iafrate~(KLI) approximation, 2D-LSDA exchange, and some of extensively used 2D-GGA 
exchange functionals. All the calculations are performed self-consistently.

\section{Density Matrix Expansion based Exchange Hole model}
The recently proposed semilocal exchange energy functionals based on the DME of the Hartree-Fock exchange 
gained more attention for studying the low dimensional quantum systems. Not only, it is one of the best 
strategies to construct an analytic expression for the exchange energy functional in 3D~\cite{kos,tsuneda}
but also in 2D~\cite{sj} as well. In 2D, the Hartree-Fock exchange energy in terms of density and exchange 
hole is given by,
\begin{equation}
 E_x=\frac{1}{2}\int d^2 r\int d^2 r'~
 \frac{\rho(\mathbf{r}) \rho_x(\mathbf{r},\mathbf{r'})}{|\mathbf{r}-\mathbf{r'}|},
 \label{eq1}
\end{equation}
where $\rho_x(\mathbf{r},\mathbf{r'})$ be the cylindrically averaged exchange hole density. It can be expressed 
in terms of the $1^{st}$ order reduced density matrix $\gamma_1(\mathbf{r},\mathbf{r'})$ as
\begin{equation}
 \rho_x(\mathbf{r},\mathbf{r'})=-\frac{\langle|\gamma_1(\mathbf{r},\mathbf{r'})|^2\rangle}{2\rho(\mathbf{r})}
 \label{eq2}
\end{equation}
with
\begin{equation}
 \gamma_1(\mathbf{r},\mathbf{r'})=2\sum_{i=1}^{occu}\phi_i^*(r)\phi_i(r'),
 \label{eq3}
\end{equation}
where $\phi_i$ are the occupied KS orbitals. On using the general coordinate transformation $(\mathbf{r},\mathbf{r}')
\rightarrow~(\mathbf{r}_{\lambda},\mathbf{u})$ with $\mathbf{r}_{\lambda}=\lambda\mathbf{r}+(1-\lambda)\mathbf{r}'$ 
and $\mathbf{u}=\mathbf{r}'-\mathbf{r}$. Here, $\lambda$ is a real number between $1/2$ and $1$ (i.e. exchange hole 
varies between maximally localized  and conventional one). Now, due to the above coordinate transformation, the 
exchange energy expression i.e. Eq.(\ref{eq1}) reduces to
\begin{equation}
 E_x = \frac{1}{2}\int d^2 r_{\lambda}\int d^2 u~\frac{\rho(\mathbf{r}_{\lambda}) \rho_x 
 (\mathbf{r}_{\lambda},\mathbf{u})}{\mathbf{u}}~,
 \label{eq4}
\end{equation}
and Eq.~(\ref{eq2}) can be rewritten as
\begin{equation}
 \rho_x (\mathbf{r}_{\lambda},\mathbf{u}) = -\frac{\langle|\gamma^t_1(\mathbf{r}_{\lambda}-(1-\lambda)\mathbf{u},
 \mathbf{r}_{\lambda} + \mathbf{u})|^2\rangle}{2\rho(\mathbf{r}_{\lambda})}~,
 \label{eq5}
\end{equation}
where $\gamma^t_1(\mathbf{r}_{\lambda}-(1-\lambda)\mathbf{u},\mathbf{r}_{\lambda}+\mathbf{u})$ be the transformed 
single-particle density matrix. Now, expanding the density matrix about $u=0$ and replacing the exponential term 
arising in it with the Bessel and Hypergeometric functions~\cite{sj}, the transformed density matrix reduces to
\begin{eqnarray}
 \varGamma_{1t}=2\rho\frac{J_1(ku)}{ku}+~~~~~~~~~~~~~~~~~~~~~~~~~~~~~~~~~~~~~~~~~~~~~~~~~\nonumber\\
 \frac{6J_3(ku)}{k^3u}\Big[4\cos^2\phi\{(\lambda^2-\lambda+\frac{1}{2})\nabla^2\rho-2\tau\} + k^2\rho\Big].\nonumber\\
 \label{eq6}
\end{eqnarray}

The choice of the expansion is comprehensible since the first term recovers the exact LDA for the homogeneous systems 
in 2D. The additional terms present besides the zeroth order LDA term takes care of the inhomogeneity involved in the 
system. Now following similar procedure as ~\cite{vsxc98} for 2D, the cylindrically averaged exchange hole from the 
DME expression Eq.(\ref{eq6}) is given by,
\begin{eqnarray}
  \rho_x(r,u)=-\frac{2J_1^2(ku)\rho}{k^2u^2}-\frac{12J_1(ku)J_3(ku)}{k^4u^2}
  \mathcal{A}\nonumber\\
-\frac{18J_3^2(ku)}{k^6u^2\rho}\mathcal{A}^2,~~~~~~
 \label{eq7}
\end{eqnarray}
where, $\mathcal{A}=2(\lambda^2-\lambda+\frac{1}{2})\nabla^2\rho-4\tau+k^2\rho$. Now, in order to further achieve 
the reliable accuracy of the newly constructed semilocal exchange functional, the expansion up to $4^{th}$ order 
in `$u$' is taken into consideration. It is noteworthy to mention that the first term in the Eq.(\ref{eq7}) corresponds 
to the exchange hole for systems with uniform electronic density. So, the coordinate transformation involved here 
is responsible for including the inhomogeneity effects but keeps the homogeneous term unaffected. It's because the 
homogeneity of the system is translationally-invariant. Therefore, we have $\lambda$ dependency appearing only in
the higher order terms (i.e. the $2^{nd}$ and $3^{rd}$ terms).

\section{Exchange energy functionals}
\label{sec3}
Now, for constructing the desired semilocal exchange functionals, the density matrix expansion and exchange hole 
model given by Eq.(\ref{eq6}) and Eq.(\ref{eq7}) are used. Here, we have replaced the laplacian term involved with 
help of the integration by parts. Thus, using these ideas and plugging Eq.(\ref{eq7}) back in Eq.(\ref{eq1}), the 
exchange energy functional becomes,
\begin{equation}
 E_x=-\int d^2r\Big[\frac{8\rho^2}{3k}+\frac{16\rho^3}{15k^3}\mathcal{B}+\frac{32\rho^4}{35k^5}
 \mathcal{B}^2\Big],
        \label{eq8}
\end{equation}
where
\begin{equation}
 \mathcal{B} = \Big(\lambda^2-\lambda+\frac{1}{2}\Big)x^2-\Big(\frac{4\tau-k^2\rho}{\rho^2}\Big)
 \label{eq9}
\end{equation}
and $x=\frac{|\nabla\rho|}{\rho^{3/2}}$ , be the dimensionless quantity called the reduced density gradient in 2D. 
Now, the newly constructed exchange energy functional given above depends on $\rho,~ \tau,~ \lambda$ and momentum 
`$k$'. The first and obvious choice of `$k$' is the Fermi momentum. Upon replacing $k_F=\sqrt{2\pi\rho}$ and $\tau$ 
by $\tau^{unif}=\pi\rho^2/2$ the homogeneous limit of the above expansion automatically can be recovered. This is 
what makes DME very special than other exchange hole models. But, instead of considering $k=k_F$, different physically 
motivated choices for $k$ can play a very crucial role in designing the exchange functional, which is the main search 
of this present work. Using spin scaling relation of the exchange energy, i.e.,
\begin{equation}
 E_x[\rho_{\uparrow},\rho_{\downarrow}]=\frac{1}{2}E_x[2\rho_{\uparrow}]+\frac{1}{2}E_x[2\rho_{\downarrow}],
 \label{eq10}
\end{equation}
the spin-polarized exchange energy functional corresponding to Eq.(\ref{eq8}) becomes,
\begin{eqnarray}
 E_x=-\sum_{\sigma=\uparrow,\downarrow}\int d^2r~\Big[\frac{32\rho_{\sigma}^2}{3k_{\sigma}}+
 \frac{128\rho_{\sigma}^3}{15k_{\sigma}^3}\mathcal{G}_{\sigma}(x_{\sigma},z_{\sigma})\nonumber\\
      +\frac{512\rho_{\sigma}^4}{35k_{\sigma}^5}\mathcal{G}^2_{\sigma}(x_{\sigma},z_{\sigma})\Big],
      \label{eq11}
\end{eqnarray}
where
\begin{equation}
 \mathcal{G}_{\sigma}(x_{\sigma},z_{\sigma})=(\lambda^2-\lambda+\frac{1}{2})\frac{x^2_{\sigma}}{2}-z_{\sigma}
 \label{eq12}
\end{equation}
``$z_{\sigma}=\frac{\tau}{\rho_{\sigma}^2}-2\pi$'' is a dimensionless quantity. The functional form of the momentum 
present in the denominator of all these terms is not unique. Only one constraint should be taken care for the momentum 
i.e. it should have the dimension of length inverse. In 3D, some forms of momentum are proposed ~\cite{kos,vs97,vsxc98}. 
The prime and transparent choice for `$k$' is `$k_F$' and the exchange functional using this becomes,
\begin{eqnarray}
  E_x=-\sum_{\sigma=\uparrow,\downarrow}\int d^2r~\frac{32\rho_{\sigma}^2}{3k_F}
  \Big[1+\frac{4\rho_{\sigma}\mathcal{G}_{\sigma}(x_{\sigma},z_{\sigma})}{5k_F^2}\nonumber\\
      +\frac{48\rho^2_{\sigma}\mathcal{G}^2_{\sigma}(x_{\sigma},z_{\sigma})}{35k_F^4}\Big].
      \label{eq13}
\end{eqnarray}
However, here we are interested in adding non-uniformity of the electronic density by making various choices for 
momentum. We have used different choices of `$k$' other than `$k_F$' and based on the physically relevant choices 
of `$k$', different exchange energy functionals are developed. So, in order to add the inhomogeneity of the system 
to the functional through momentum, one needs to add suitable terms intuitively having density dependency, which 
upon imposing the homogeneity limit should correctly recover the LDA exchange functional. In principle, this 
happens because the new exchange energy functional i.e., Eq.(\ref{eq13}) depends on dimensionless quantities 
$x_{\sigma}$ and $z_{\sigma}$ as for homogeneous systems, $x_{\sigma}$ becomes zero as it depends on $\nabla\rho$. 
Similarly, $z_{\sigma}$ goes to zero when $\tau=\tau^{unif}$(HEG). Using these ingredients, functionals are developed 
in the next subsections.

\subsection{Density gradient dependent momentum}
We have added the dimensionless reduced density gradient based terms to the Fermi momentum. The addition of 
$x_{\sigma}^2$ to $k_F$, obeys all the conditions i.e. new momentum $\overline{k}_{F,g}$ has the dimension of 
length inverse and becomes $k_F$ in the uniform density limit. We have proposed the first modification to the 
Fermi momentum through
\begin{equation}
 \overline{k}_{F,g}=k_F(1+\alpha x_{\sigma}^2),
 \label{eq15}
\end{equation}
where $\alpha$ is an adjustable parameter which takes care of the gradient effect.
Using $\overline{k}_{F,g}$ from Eq.~(\ref{eq15}), in Eq.~(\ref{eq13}) 
the new semilocal exchange energy functional $E_x^{GDM}$~(exchange energy with gradient 
dependent momentum) becomes,
\begin{eqnarray}
E_x^{GDM}[x_{\sigma},z_{\sigma}]&&=-\sum_{\sigma=\uparrow,\downarrow}\int d^2r
 \Big[\frac{32\rho_{\sigma}^2}{3k_{F,g}}+\nonumber\\
 &&A\frac{128\rho_{\sigma}\mathcal{G}_{\sigma}(x_{\sigma},z_{\sigma})}{15\overline k_{F,g}^3}
      +B\frac{512\rho^2_{\sigma}\mathcal{G}^2_{\sigma}(x_{\sigma},z_{\sigma})}{35\overline k_{F,g}^5}\Big].\nonumber\\~~~~
      \label{eq16}
\end{eqnarray}
Since, in the present study, the density matrix expansion is terminated at the $2^{nd}$ order. As a matter of which, 
the exchange hole is not exact. To take care the above fact, we have introduced two adjustable parameters `$A$' and 
`$B$' which will be fixed later. Also, the first term within square bracket is different from the LSDA exchange energy
because of the presence of $\overline{k}_{F,g}$ in the denominator. For this parameterization of higher order terms are 
necessary.

\subsection{Kinetic energy dependent momentum}
Our next attempt is to construct the semilocal exchange functional through K.E. dependent momentum.
Since, the term $z_{\sigma}$ present in the exchange energy functional expression i.e., Eq.(\ref{eq13}), 
contains the kinetic energy density as one of its main ingredients. Hence, the inclusion of such terms 
in $k_F$, makes the momentum kinetic energy density dependent. In this way, we have included the non-uniformity  
through momentum by means of $\tau$. It is conspicuous that addition of some fraction of $z_{\sigma}$ 
to $k_F$ obeys the dimension and uniform density limit restrictions. Thus, a new form of the transformed 
momentum is proposed to be,
\begin{equation}
 \overline{k}_{F,t} =k_F(1+\alpha  z_{\sigma}),
 \label{eq17}
\end{equation}
where $\alpha$ be an adjustable parameter introduced to add a small fraction of the inhomogeneity through $z_{\sigma}$. 
Now, upon substituting the changed momentum from Eq.(\ref{eq17}) in the exchange expression of Eq.(\ref{eq13}), readily 
leads to the following exchange energy functional $E_x^{TDM}$~($\tau$ dependent momentum) similar to  Eq.(\ref{eq16}) 
having the form, 
\begin{eqnarray}
 E_x^{TDM}[x_{\sigma},z_{\sigma}]&&=-\sum_{\sigma=\uparrow,\downarrow}\int d^2r
 \Big[\frac{32\rho_{\sigma}^2}{3k_{F,t}}+\nonumber\\
 &&A\frac{128\rho_{\sigma}\mathcal{G}_{\sigma}(x_{\sigma},z_{\sigma})}{15\overline k_{F,t}^3}
      +B\frac{512\rho^2_{\sigma}\mathcal{G}^2_{\sigma}(x_{\sigma},z_{\sigma})}{35\overline k_{F,t}^5}\Big].\nonumber\\~~~~
      \label{eq18}
\end{eqnarray}
where $A$ and $B$ are adjustable parameters similar to that involved in Eq.(\ref{eq16}). So, Eq.(\ref{eq18}) only
differs from Eq.(\ref{eq16}) by a different choice of momentum i.e. $\overline{k}_{F,g}$ is replaced by 
$\overline{k}_{F,t}$.

\begin{table}
\begin{center}
\caption{Tabulated are the adjusted values of all the constants present in the exchange energy functionals 
$E_x^{GDM}$, $E_x^{TDM}$, and $E_x^{GTDM}$}
\begin{tabular}{c  c  c  c }
\hline\hline
Functional   ~~~&$\alpha$ ~~~&$A$  ~~~&$B$   \\ \hline
$E_x^{GDM}$  ~~~& 0.001  ~~~&0.1  ~~~&0.3951 \\
$E_x^{TDM}$  ~~~& 0.001   ~~~&0.1  ~~~&0.0946 \\
$E_x^{GTDM}$ ~~~& 0.001  ~~~&0.1  ~~~&0.442 \\
\hline\hline
\label{Table1}
\end{tabular}
\end{center}
\end{table}

\begin{figure}
\begin{center}
\includegraphics[width=3.6in,height=2.6in,angle=0.0]{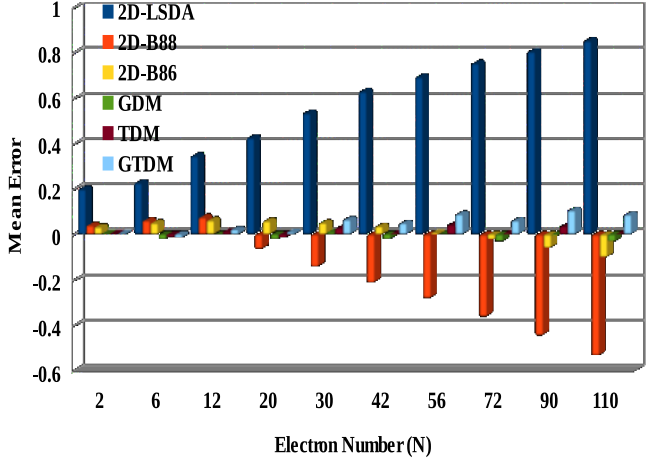} 
\end{center}
\caption{Shown are the mean error of exchange energies for few electron parabolic quantum dots plotted against 
the electron number.}
\label{fig1}
\end{figure}
\subsection{Reduced density gradient and kinetic energy density dependent momentum}
So far we have used $x_{\sigma}$ and $z_{\sigma}$ individually, to enhance the inhomogeneity effects in the semilocal 
functionals through the momentum. It is indicative that, use of both $x_{\sigma}$ and $z_{\sigma}$ together is a 
noticeable option because the addition of both terms in appropriate order satisfies the inverse length dimension of 
momentum and the modified momentum recovers the Fermi momentum for uniform densities. The simplest form of both 
kinetic energy density and reduced density gradient dependent momentum is,
\begin{equation}
 \overline{k}_{F,g,t}=k_F\{1+\alpha(x_{\sigma}^2+z_{\sigma})\},
 \label{eq19}
\end{equation}
where $\alpha$ is an adjustable parameter defined in the same manner as Eq.(\ref{eq15}) and Eq.(\ref{eq17}). The 
exchange energy functional $E_x^{GTDM}$(gradient and $\tau$ dependent momentum) with $\overline{k}_{F,g,t}$ is,
\begin{eqnarray}
  E_x^{GTDM}[x_{\sigma},z_{\sigma}]&&=-\sum_{\sigma=\uparrow,\downarrow}\int d^2r
 \Big[\frac{32\rho_{\sigma}^2}{3k_{F,g,t}}+\nonumber\\
 &&A\frac{128\rho_{\sigma}\mathcal{G}_{\sigma}(x_{\sigma},z_{\sigma})}{15\overline k_{F,g,t}^3}
      +B\frac{512\rho^2_{\sigma}\mathcal{G}^2_{\sigma}(x_{\sigma},z_{\sigma})}{35\overline k_{F,g,t}^5}\Big],\nonumber\\~~~~~~
      \label{eq20}
\end{eqnarray}
where `$A$' and `$B$' are again tunable constants. The effects of higher-order terms in the density matrix expansion 
can be included through these parameters.

\begin{table*}[ht]
\begin{center}
\caption{Shown in the table are the exchange energies (in atomic units) for parabolic QDs calculated using the 
new 2D exchange energy functionals $-E_x^{GDM}$, $-E_x^{TDM}$, and $-E_x^{GTDM}$. The $1^{st}$ column is for the 
number of electrons `$N$', $2^{nd}$ is for confinement strength $\omega$. The self-consistent calculations for 
2D-EXX~(KLI), 2D-LDA, 2D modified B86, 2D-B88 are shown in succeeding columns. The last three columns are 
the results for our constructed functionals. The last row contains the MAPE ($\Delta$). }
\begin{tabular}{c  c  c  c  c  c  c c c}
\hline\hline
N  &$\omega$&$-E_x^{2D-EXX}$&$-E_{x}^{2D-LDA}$&$-E_x^{2D-B86}$&$-E_x^{2D-B88}$ &$-E_x^{2D-GDM}$&$-E_x^{2D-TDM}$&$-E_x^{2D-GTDM}$\\ \hline
2  & 1/6    &0.380          &0.337            &0.368          &0.364           & 0.388        &0.388          &0.388\\
2  & 0.25   &0.485          &0.431            & 0.470         &0.464           & 0.493       &0.494      &0.494 \\
2  & 0.50   &0.729          &0.649            & 0.707         &0.699           &0.738       &0.738      &0.738 \\
2  & 1.00   & 1.083         &0.967            & 1.051         &1.039           & 1.089       &1.089     &1.089\\
2  & 1.50   &1.358          &1.214            & 1.319         &1.344           & 1.358       &1.358      & 1.358\\
2  & 2.50   &1.797          &1.610            & 1.748         &1.728           &1.777       &1.777      &1.777\\
2  & 3.50   &2.157          &1.934            & 2.097         &2.074           & 2.106       &2.107      &2.106\\
6  & 0.25   &1.618          &1.531            &1.603          &1.594           &1.646       &1.639      &1.643\\
6  & 0.50   &2.470          &2.339            &2.444          &2.431           &2.504       &2.494     &2.498  \\
6  & 1.00   &3.732          &3.537            &3.690          &3.742           &3.769       &3.756     &3.761 \\
6  & 1.50   &4.726          &4.482            &4.672          &4.648           &4.763       &4.748     &4.752 \\
6  & 2.50   &6.331          &6.008            &6.258          &6.226           &6.356       &6.338     &6.341 \\
6  & 3.50   &7.651          &7.264            &7.562          &7.525           &7.644       &7.633      &7.631\\
12 & 0.50   &5.431          &5.257            &5.406          &5.387           & 5.470       &5.462     &5.458 \\
12 & 1.00   &8.275          &8.013            & 8.230         &8.311           &8.312       &8.302     &8.293\\
12 & 1.50   &10.535         &10.206           &10.476         &10.444          &10.562      &10.550    &10.538 \\
12 & 2.50   &14.204         &13.765           &14.122         &14.080          &14.19      &14.172    &14.158  \\
12 & 3.50   &17.237         &16.709           &17.136         &17.165          & 17.098      & 17.138    &17.129  \\
20 & 0.50   &9.765          &9.553            &9.746          &9.819           &  9.788      &9.810      &9.801 \\
20 & 1.00   &14.957         &14.638           &14.919         &15.014          &14.970       &15.002    &14.986\\
20 & 1.50   &19.108         &18.704           &19.053         &19.159          & 19.102      &19.143    &19.122\\
20 & 2.50   &25.875         &25.334           &25.796         &25.973          & 25.888      &25.871     &25.843  \\
20 & 3.50   &31.491         &30.837           &31.392         &31.603          & 31.433      & 31.437    &31.392\\
30 & 1.00   &23.979         &23.610           &23.953         &24.091          &24.041      &24.029    &24.003  \\
30 & 1.50   &30.707         &30.237           &30.665         &30.836          & 30.753      &30.738    &30.704 \\
30 & 3.50   &50.882         &50.115           &50.794         &51.068          & 50.739      &50.721     & 50.667\\
42 & 1.00   &35.513         &35.099           &35.503         &35.671          & 35.596      &35.583    & 35.548\\
42 & 1.50   &45.560         &45.032           &45.538         &45.747          & 45.634      &45.617   & 45.573\\
42 & 2.50   &62.051         &61.339           &62.007         &62.286          & 62.059      &62.044     &61.986 \\
42 & 3.50   &75.814         &74.946           &75.748         &76.085          & 75.734      & 75.677    & 75.634\\
56 & 1.00   &49.710         &49.256           &49.722         &49.919          & 49.802      &49.789    &49.743\\
56 & 1.50   &63.869         &63.289           &63.871         &64.117          &63.939      & 63.921    &63.866\\
56 & 2.50   &87.164         &86.378           &87.148         &87.479          &87.146      &87.108     & 87.047\\
56 & 3.50   &106.639        &105.684          &106.609        &107.010         &106.489     &106.401    & 106.369 \\
72 & 1.00   &66.708         &66.219           &66.746         &66.972          &66.823      &66.810    & 66.755\\
72 & 1.50   &85.814         &85.186           &85.844         &86.129          &85.913      &85.895    &85.829 \\
72 & 2.50   &117.312        &116.456          &117.327        &117.712         &117.339     &117.305    & 117.232\\
72 & 3.50   &143.696        &142.650          &143.697        &144.163         &143.606     &143.5   & 143.469\\
90 & 1.00   &86.631         &86.111           &86.698         &86.954          &86.756      &86.743     &86.679 \\
90 & 1.50   &111.558        &110.889          &111.622        &111.946         &111.659     &111.639    & 111.561\\
90 & 2.50   &152.723        &151.808          &152.779        &153.217         &152.699     &152.686   &152.587\\
90 & 3.50   &187.262        &186.139          &187.306        &187.838         &187.079     &186.960    &186.922\\
110& 1.00   &109.595        &109.048          &109.695        &109.981         &109.748     &109.734   &109.661\\
110& 1.50   &141.255        &140.548          &141.357        &141.720         &141.386     &141.365    &141.277 \\
110& 2.50   &193.617        &192.647          &193.715        &194.210         &193.6231     &193.615    &193.496 \\
110& 3.50   &237.612        &236.420          &237.706        &238.306         &237.474     & 237.356   &237.297\\
\hline
$\Delta$&   &               & 4.2             &0.71           & 1.04        &{\bf{0.44}}  &{\bf{0.40}}&{\bf{0.40}}\\
\hline\hline
\label{Table2}
\end{tabular}
\end{center}
\end{table*}

\section{The correlation energy functional}
Now, we will try to construct the correlation energy functional compatible with the three semilocal exchange 
energy functionals constructed above. The 2D-LDA correlation energy functional~\cite{attaccalite} is a commonly 
used functional in the calculations of 2D quantum systems. This functional was constructed by the interpolation 
of the low-density limit from Diffusion Monte Carlo~(DMC) data and high-density limit from 2D-LDA exchange-
correlation energy functional~\cite{rk} having a parameterized form, 
\begin{eqnarray}
 \epsilon_c^{LDA}(r_s,\zeta)=(e^{-\beta r_s}-1)\epsilon_x^{(6)}(r_s,\zeta)+\nonumber ~~~~~~~~~~~\\
 \alpha_0(r_s) +\alpha_1(r_s)\zeta^2+ \alpha_2(r_s)\zeta^4,
 \label{eq21}
\end{eqnarray}
where $r_s=1/\sqrt{\pi\rho}$, $\zeta$ be the usual spin-polarization and $\epsilon_x^{(6)}(r_s,\zeta) = 
\epsilon_x(r_s,\zeta)-(1+\frac{3}{8}\zeta^2+\frac{3}{128}\zeta^4)\epsilon_x(r_s,0)$ be the Taylor expansion of
$\epsilon_x$ beyond fourth order in $\zeta$. Here, the 2D-LSDA exchange energy term, $\epsilon_x = -2\sqrt{2}
[(1+\zeta)^{3/2}+(1-\zeta)^{3/2}]/3\pi r_s$. The functional form of $\alpha_i(r_s)$ is taken as a 2D generalization 
form from electron-gas correlation of Perdew-Wang~\cite{perdew-wang} and is given by,
\begin{eqnarray}
 \alpha_i(r_s)= A_i+\Big(B_i+C_ir_s^2+D_ir_s^3\Big)\nonumber ~~~~~~~~~~~~~~~~~~~~~~~~~\\
 \times ln\Big (1+\frac{1}{E_ir_s+F_ir_s^{3/2}+G_ir_s^2+H_ir_s^3}\Big).~~~~
 \label{eq22}
\end{eqnarray}
The values of all the constants present in the above Eq.(\ref{eq22}) are given in the Table II of reference
\cite{attaccalite}. This correlation functional depends on spin-polarization $\zeta$ and electron density 
$\rho$ via $r_s$, which make this functional local. However, the application of this LDA functional to the 
parabolic QDs overestimates the correlation energy up to a large extent which can be observed from the 
Table~\ref{Table3}. So, in order to apply this correlation functional to non-uniform systems, modifications 
to the correlation functional is desirable. Thus, we have proposed that the non-local effects of the real 
system can be engineered into the LDA correlation energy via a parametric form of our exchange energy functional
constructed above. In fact, use of exchange energy enhancement factor in the correlation energy is encountered 
in previous studies of 3D correlation functionals~\cite{vsxc98,becke-cor-98}. Analogous to 3D, here, we have
used a modified form of $E_x^{GTDM}$ with some convenient parameters. The modified form of momentum from 
Eq.(\ref{eq19}) with a different constant can be written as,
\begin{equation}
 \overline{k}_{F,g,t}=k_F\{1+\delta(x_{\sigma}^2+z_{\sigma})\}=k_F\Gamma_{\sigma}(x_{\sigma},z_{\sigma}).
 \label{eq23}
\end{equation}
Now, using the above form of momentum from Eq.(\ref{eq23}) in place of $k_{\sigma}$ present in the Eq.(\ref{eq11}), 
one can easily rewrite Eq.~(\ref{eq11}) as,
\begin{eqnarray}
  E_x=-\sum_{\sigma=\uparrow,\downarrow}\int d^2r~\frac{32\rho_{\sigma}^2}{3k_F}
  \Big[\frac{L}{\Gamma_{\sigma}}+
  \frac{M~4\rho_{\sigma}\mathcal{G}_{\sigma}(x_{\sigma},z_{\sigma})}
  {5k_F^2 \Gamma_{\sigma}^3}\nonumber\\
      +\frac{N~48\rho^2_{\sigma}\mathcal{G}^2_{\sigma}(x_{\sigma},z_{\sigma})}
      {35k_F^4\Gamma_{\sigma}^5}\Big],~~~
      \label{eq24}
\end{eqnarray}
where L, M, and N are parameters introduced to account the effects of neglected higher order terms. The above 
Eq.(\ref{eq24}) is written in a similar way as $E_x^{GTDM}$ but with a small modification in the enhancement 
factor. This modification is necessary to give proper multiplicative factor to the correlation functional. In 
this section, we are not interested in calculating exchange energy functional Eq.(\ref{eq24}). But to incorporate
the non-local effects in the correlation functional, multiplication by enhancement factor like term present 
within square bracket in the above Eq.(\ref{eq24}) is desirable. We denote this term as `$\mathbf{f_{\sigma}}$'.
The term `$\mathbf{f_{\sigma}}$ ' is a dimensionless quantity and will not change the dimension of any quantity 
when it will be multiplied by the same. Hence, taking the local contribution from Eq.(\ref{eq21}) and inducing 
non-uniformity via `$\mathbf{f_{\sigma}}$', we have proposed a new spin-polarized correlation energy functional 
to be,
\begin{equation}
 E_{c,\sigma}^{NIL}=\sum_{\sigma=\uparrow,\downarrow}\int d^2r~\epsilon_{c,\sigma}^{LDA}(r_s,\zeta)
 \mathbf{f_{\sigma}}(x_{\sigma},z_{\sigma})~.
 \label{eq25}
\end{equation}
The above non-local effect induced LDA correlation functional $E_{c,\sigma}^{NIL}$ will be completed when 
appropriate values for all the parameters will be defined successfully. This task will be completed in the 
next section by comparing the result with the exact values for the parabolic quantum dots.

\section{Results and Discussion}
For numerical calculations, we have applied the newly proposed functionals to parabolically confined quantum 
dots~(QD) which are also known as artificial atoms. The external potential for the QD is $\omega^2r^2/2$ with 
confinement strength $\omega$. The exact exchange~(EXX) results for the QDs are calculated using optimized 
potential method~(OPM) with Krieger-Li-Iafrate~(KLI)~\cite{kli} approximation. The OCTOPUS~\cite{octopus} 
code is used  for all the self-consistent calculations and the outputs such as spin-polarized density and 
spin-polarized kinetic energy densities of EXX are used as input for the newly constructed functionals. To 
compare the results of exchange energy functionals, some of the previously constructed exchange energy 
functionals like 2D-LDA~\cite{rk}, 2D-B88~\cite{vrmp}, and 2D-modified GGA~\cite{prvm} are also calculated 
self-consistently using OCTOPUS. All the exchange energy functionals are analyzed, by varying the number 
of electrons from 2 to 110 and the confinement strength from 0.25 to 3.5 in the parabolic quantum dot.

Here, our first task is to give appropriate values to all the constants present in the exchange and correlation 
energy functionals. All the three exchange energy functionals $E_x^{GDM}$, $E_x^{TDM}$, and $E_x^{GTDM}$ contain 
four adjustable parameters such as $A, B, \lambda,$ and $\alpha$. The constant $\lambda$ was defined for the 
localization of the exchange hole and for the present calculations, we have chosen maximally localized exchange 
hole, which implies $\lambda=0.5$. The first term in all these exchange functionals is LSDA like term. Hence, all the 
succeeding terms should be corrections to the LSDA and the effects should decrease gradually. So, keeping this 
in mind and comparing results for two electrons parabolic QD with EXX-KLI, we have fixed the value of `$A$'. 
Now, we are left with two more parameters $\alpha$ and $B$. The parameter `$\alpha$' is the constant introduced 
to take care the addition of inhomogeneity to $k_F$ and `B' is multiplied to the coefficient of $3^{rd}$ term 
present within all the square brackets of exchange energy functionals. Thus, the choice of $\alpha$ and $B$ are 
interdependent. We select $\alpha$ as a very small real number. This will add a little non-uniformity to $k_F$. 
Because for a higher value of $\alpha$ the result will exceed from exact values. we have calculated exchange 
energies for two electron parabolic quantum dot varying `$\omega$' from 0.25 to 3.5 and for each system, we have
determined `$B$' which gave zero error. Finally, the mean value of all $B$'s is taken as concluding value of it. 
All these procedures are followed to fix the parameters of exchange energy functionals and values for these 
parameters are given in Table~\ref{Table1}. The settled values for all the adjustable parameters depend on QD systems.

The Table~\ref{Table2} comprises of the number of electrons~($N$), confinement strength~($\omega$), and four 
known functionals, in the first six columns. The last three columns are results obtained from the newly 
constructed exchange energy functionals. The mean absolute percentage of error~(MAPE) for all the functionals 
are given for comparison. The competitive and better performance of the new functionals to calculate the 
exchange energies are clear from the Table~\ref{Table2}. Exchange energy functional $E_x^{GDM}$, $E_x^{TDM}$, 
and $E_x^{GTDM}$ give MAPE as 0.44, 0.40, and 0.40 respectively for all 46 calculations. Figure~\ref{fig1}, is
the mean error~(ME) plot for all the discussed functionals. The heights of different color bars represent the 
ME corresponding to the exchange energy functionals considered in the present study.

\begin{table*}[ht]
\begin{center}
\caption{The first two columns represent the number of electrons~($N$) and the confinement strength $(\omega)$.
All the references for exact values of total energies of different quantum dots are given below the table. The 
last row contains the MAPE ($\Delta$).}
\begin{tabular}{c  c  c  c  c  c  c  c  c  c}
\hline\hline
N &$\omega$  &$E_{tot}^{ref^*}$&$E_{tot}^{2D-EXX}$&$-E_c^{ref}$&$-E_c^{2D-LDA}$&$-E_c^{NIL}$&$-E_{xc}^{ref^{\dagger}}$&$-E_{XC}^{LDA}$&-$E_{XC}^{mod}$\\ \hline
2 &1/6       &$2/3^a$          &0.7686            & 0.1020     &0.1221         &0.1055      &0.4936         &0.4721         &0.4935\\
2 &0.25      &$0.9324^b$       &1.0462            & 0.1138     &0.1390         &0.1193      &0.5987         &0.5819         &0.6133\\
2 &1.00      &$3^a$            &3.1619            & 0.1619     &0.1987         &0.1641      &1.246          &1.1737         &1.2531\\
6 &0.25      &$6.995^b$        &7.3910            & 0.3960     &0.4574         &0.3924      &2.014          &2.0112         &2.0314\\
6 &$1/1.89^2$&$7.6001^c$       &8.0211            & 0.4210     &0.4732         &0.4054      &2.156          &2.1372         &2.1614\\
6 &0.42168   &$10.37^d$        &10.8204           & 0.4504     &0.5305         &0.4524      &2.6784         &2.6604         &2.7034\\  \hline

$\Delta$& $-$   &  $-$               &  $-$                & $-$           &18.37          &{\bf 2.46}  &  $ -$           &2.44           &{\bf 0.84}\\
\hline\hline
\label{Table3}
\end{tabular}
\begin{flushleft}
$*$  All the reference results are discussed in References~\cite{prm08,rpp10}  \\
$\dagger~E_{xc}^{ref} = E_x^{2D-EXX}+E_c^{ref}$\\
a-Analytic solution by Taut from Ref.~\cite{taut}.\\
b-CI data from Ref.~\cite{rcbg06}.\\
c-Diffusion QMC data from Ref.~\cite{pul03}.\\
d-Variational QMC data from Ref.~\cite{srshpn03}.\\
\end{flushleft}
\end{center}
\end{table*}

The correlation functional $E_{c,\sigma}^{NIL}$ from Eq.(\ref{eq25}) contains four parameters $L$, $M$, $N$, 
and $\delta$, to be fixed by comparing the correlation energy with the exact results. In DFT correlation 
energy can be written as $E_c^{ref}=E_{tot}^{ref}-E_{tot}^{EXX}$, where $E_{tot}^{ref}$ is the exact total 
energy of the system and $E_{tot}^{EXX}$ is the total energy of the system taking EXX without any correlation. 
By analyzing the correlation energy of the parabolic QD with $2$ electrons and $\omega=1$, we have fixed the 
constants $\delta=0.01$, $L=0.8825$, and $M=0.1$. The value of $N$ is taken as the mean of all $N$'s that gives 
zero difference between the exact and calculated values for a set of parabolic quantum dots. We have considered 
the first two closed shell parabolic QDs. The correlation energy values for two and six electrons are given in 
Table~\ref{Table3}. For comparison, we have given the exact reference values and 2D-LDA correlation~\cite{attaccalite}. 
In addition to these results, we have also given the combined exchange and correlation energy $E_{xc}$ results 
for the same set of QDs in Table~\ref{Table3}. We have combined $E_x^{TDM}$ and $E_{c,\sigma}^{NIL}$ for the 
preliminary testing and exchange-correlation energy for $E_{xc}^{mod}=E_x^{TDM}+E_{c,\sigma}^{NIL}$ are given 
in the Table~\ref{Table3}. We have considered only one functional $E_x^{TDM}$ here. Similar results will be obtained
for $E_x^{GDM}$ and $E_x^{GTDM}$. It is clear from Fig~\ref{fig1} that new functionals $E_x^{TDM}$ and $E_x^{GTDM}$ 
possess the positive mean error for the higher number of electrons. For higher number of electrons some of these ME will 
be compensated by negative mean error of $E_{c,\sigma}^{NIL}$. The improvement of the proposed correlation functional 
can be easily observed from the MAPE in Table~\ref{Table3}. Also the combined effect of both exchange-correlation 
energy functional $E_{xc}^{mod}$ performs well in parabolic quantum dots.

\section{Conclusions}
We have developed three semilocal exchange energy functionals based on the density matrix expansion and a 
correlation energy functional based on the modification of LDA correlation functional by one of the newly 
constructed exchange energy functional. The non-local effects are added to the functionals by modifying the 
Fermi momentum. The Fermi momentum is modified by using reduced density gradient and kinetic energy density. 
The parameters introduced in the exchange and correlation energy functionals are set by some restrictions 
and comparing the result with the exact values. In principle, a new set of parameters can be proposed, taking 
different 2D systems which will give the better result for that system. All the functionals are tested and 
analyzed for quantum dot systems with a different number of confined electrons. The newly proposed exchange 
energy functionals are believed to achieve encouraging performance in two-dimensional many-electron calculations. 
The proposed correlation energy functional excellently improves over the LDA correlation energy functional when 
applied to quantum dots.

\end{document}